\documentclass[%
 amsmath,amssymb,
 aps,
prl,twocolumn,showpacs,
floatfix
]{revtex4-1}
\usepackage{graphicx}% Include figure files
\usepackage{dcolumn}% Align table columns on decimal point
\usepackage{bm}% bold math

\begin{document}

\title{Dynamical many-body phases of the parametrically driven, dissipative Dicke model}

\author{R. Chitra}
 \affiliation{Institute for Theoretical Physics, ETH Zurich, 8093 Z{\"u}rich, Switzerland}
\author{O. Zilberbeg}%
\affiliation{Institute for Theoretical Physics, ETH Zurich, 8093 Z{\"u}rich, Switzerland}

\begin{abstract}  The dissipative Dicke model  exhibits a fascinating out-of-equilibrium many-body phase transition as a function of a coupling between a driven photonic cavity and numerous two-level atoms. We study the effect of a time-dependent parametric modulation of this coupling, and discover a rich phase diagram as a function of the modulation strength.  We find that in addition to the established normal and super-radiant phases, a new phase with pulsed superradiance which we term  dynamical normal phase appears when the system is parametrically driven. Employing different methods, we characterize the different phases and the transitions between them. Specific  heed is paid to the role of dissipation in determining the phase boundaries. Our analysis paves the road for the experimental study of dynamically stabilized phases of interacting light and matter. 
\end{abstract}
\maketitle

Experimental progress in control and manipulation of light-matter quantum systems has generated a growing interest in many-body phenomena out of equilibrium~\cite{Carusotto2013}. Well established examples of such systems include ultracold atomic or ionic quantum gases in high finesse optical cavities~\cite{Bloch2008}, semiconductor microcavities in the strong coupling regime~\cite{kavokin2007microcavities, Carusotto2013}, and superconducting qubits in microwave resonators~\cite{wallraff2004strong,majer2007coupling}. The  engineered interplay between light and matter  in these systems  has led to the observation of a host of  fascinating collective 
phases and quantum phase transitions  including superfluidity in  polaritons~\cite{amo2009superfluidity} and the  super-radiant Dicke phase transition in
a Bose-Einstein condensate (BEC) coupled to an optical cavity~\cite{brennecke2013}. 

A defining feature of many of these systems is that they are  inherently driven and subject to dissipation. Driven dissipative systems are usually treated within a rotating frame formalism. This effectively renders the problem time-independent  with the important feature that the asymptotic steady states
are necessarily out of equilibrium.  However,  parametric driving of the system often does not allow the usual rotating frame simplifications. Consequently, the resulting interplay between interactions, dissipation, and parametric driving, could  lead to novel and exotic steady-state physics that has no counterpart in the undriven case.  Parametric driving is increasingly used as an experimental tool  in diverse contexts, e.g.~,  in the
  generation of Floquet topological insulators~\cite{Floquet},  improved measurement fidelity with squeezed quantum states~\cite{orzel2001squeezed,rugar2004single} and unconventional phenomena in cavity QED~\cite{QED}. 
  
A prime example of a system exhibiting light-matter collective phenomenon is the  Dicke model~\cite{Dimer2007}. Here, a bosonic/cavity mode is coupled to a large number of two-level atoms. It exhibits a $Z_2$ 
symmetry breaking quantum phase transition from a normal phase (NP), where all the atoms are in their ground state and the cavity is empty, to a super-radiant phase (SP), where the atoms are excited and the cavity is in a coherent state. This model has recently been realized by coupling  the external degree of freedom of  a BEC to a quantized mode of a laser-driven optical cavity, and the theoretically predicted non-equilibrium phase transition has been observed~\cite{brennecke2013, hemmerich2014}. Moreover, the inevitable photon leakage out of the cavity as well as dissipation of the BEC has been shown to lead to a considerable modification of the critical exponents of the transition~\cite{Nagy2011}. 

\begin{figure}[ht]                   
\centering
\includegraphics[width=50mm]{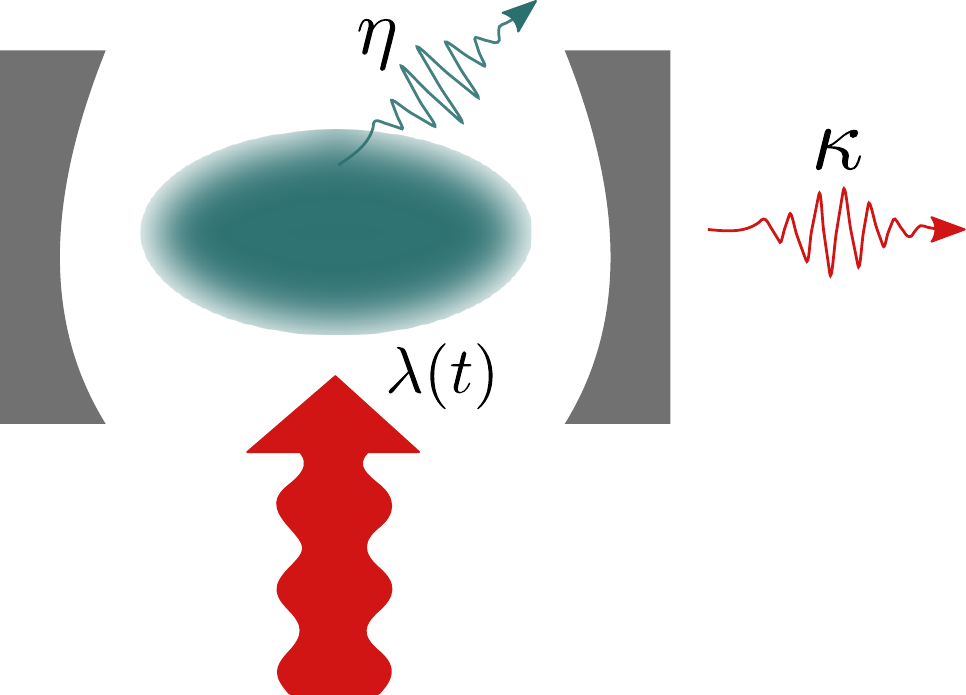}
\caption{\label{fig:setup} A sketch of a parametrically driven  Dicke  model. A single-mode cavity is driven by a laser beam with an oscillating laser-field power $P(t)$. The cavity is naturally leaky with a dissipation rate $\kappa$. Inside the cavity, an atomic cloud is cooled and forms a Bose-Einstein condensate that is coupled to the driving laser with coupling $\lambda(t)\propto \sqrt{P(t)}$ [cf.~Eq.~\eqref{dickeHam}]. The atoms are also coupled to an environment with a dissipation rate $\eta$. The system is best described by Liouvillian dynamics [cf.~Eq.~\eqref{liouvillian}]. }
\end{figure}

\begin{figure*}[ht]
\centering
\includegraphics[width=\textwidth]{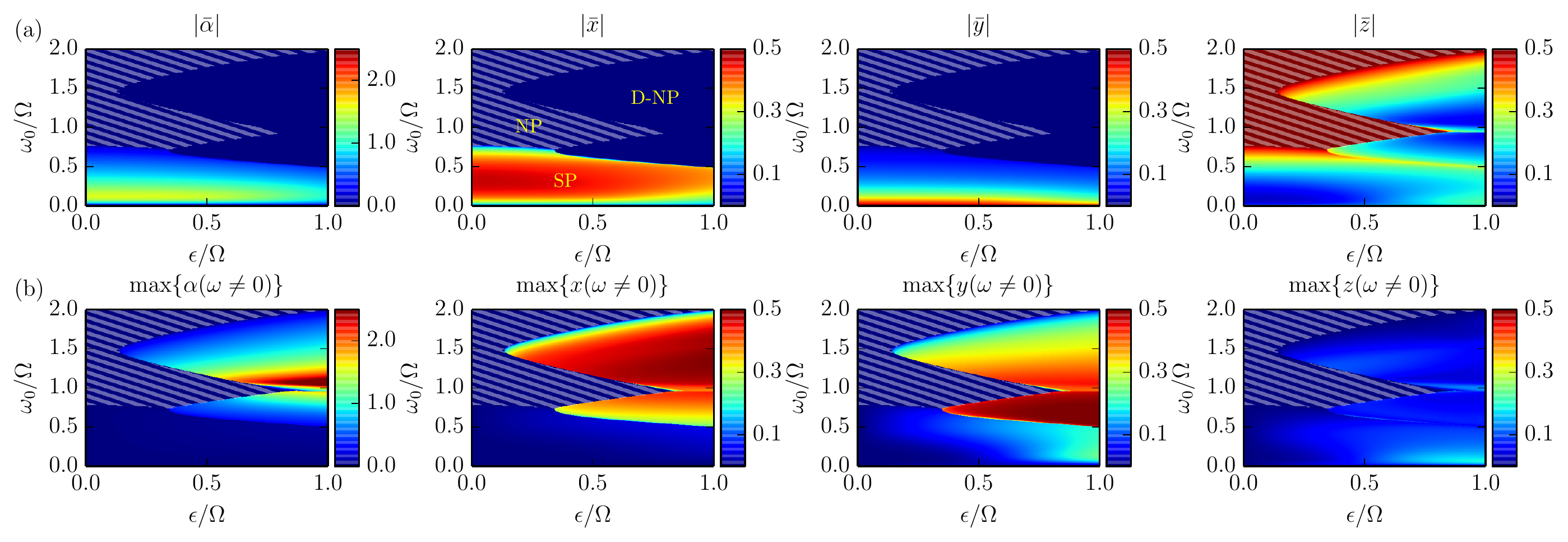} 
\caption{\label{fig:diagram} Characteristics of the dynamical phase diagram of the parametrically driven Dicke model as a function of cavity/atoms frequency $\omega_0=\omega_c=\omega_a$ and parametric modulation strength $\epsilon$ for bare coupling $\lambda_0=0.4\Omega$ and dissipation rates $\kappa=\eta=0.1\Omega$. We obtain these characteristics from the numerical steady-state solutions of the mean field equations [cf.~Eqs.~\eqref{amean}-\eqref{ymean}]. Each point, with a $10^{-3}$ resolution, is a result of a different numerical integration. Superimposed (striped overlay) is the normal mode stability zone [cf.~Eq.~\eqref{cla-eom}]. (a) Density plots of the absolute value of the steady-state time-averaged order parameters. (b) In steady-state the order parameters are oscillating with a complex beat structure~\cite{supmat}. The density plots are of the amplitude of the maximal frequency $\omega$ contributing to this oscillation. We see three distinct phases appearing as a function of $\epsilon$, i.e.~, extensions of the normal and super-radiant phases (NP and SP), as well as a novel dynamical phase (D-NP). These three phases meet at a shared multicritical point. The properties of each phase are summarized in Table \ref{table1}.}
\end{figure*}

In this work, we analyze the impact of parametric driving on the phase diagram of the dissipative Dicke model.  Specifically, we consider a modulation of the atom-cavity coupling, which is easily realizable in current experimental setups, see Fig.~\ref{fig:setup}. Using a combination of mean field theory and effective Hamiltonians, we obtain a rich phase diagram comprising: (i) the NP with parametric amplification, (ii) the SP phase, and (iii) a novel dynamical normal phase (D-NP), which appears to be a dynamically rotating NP with pulsed super-radiance. We elucidate the vital role that dissipation plays in modifying the complex phase topography of this nonequilibrium system. Our analysis presents parametric driving as a promising frontier in the search for exotic collective phases in light-matter systems.

The single mode parametrically driven Dicke model is described by the Hamiltonian
\begin{equation}
H (t) = \hbar\omega_c a^\dagger a + \hbar\omega_a \sum_{i=1}^N s_z^i + {\frac{2\hbar\lambda(t)}{\sqrt{N}}} \sum_{i=1}^N s_x^i (a + a^\dagger)\,,
\label{dickeHam}
\end{equation}
where $s^i_\alpha$  with $\alpha=x,y,z$ are the spin operators describing the $i^{\rm th}$ two level atom, and $a, a^\dagger$ represent the cavity creation and
annihilation operators. The cavity's resonance frequency is $\omega_c$, whereas the atoms are considered to be identical with level spacing $\hbar\omega_a$. We consider a time-dependent coupling between the atoms and the cavity of the form $\lambda(t)= \lambda_0 + \epsilon \cos(2 \Omega t)$. Such a coupling is easily generated by a modulation of the laser power that drives the cavity~\cite{epsilon}. 
 For $\epsilon=0$, the system exhibits the well known continuous phase transition from a NP to a SP when the coupling $\lambda_0  \ge \sqrt{\omega_c \omega_a}/2$~\cite{Dimer2007}. For $\epsilon\neq 0$, we reiterate that the parametric driving described here cannot be rotated away by a suitable choice of frame. 
Indeed, recent treatments of similar modulations of the Dicke model were addressed using a mapping to parametric oscillators, and a partial phase diagram for the NP was obtained \cite{bastidas2012, vacanti2012}. Here, we explicitly include dissipation for both the cavity and the atoms (see Fig.~\ref{fig:setup}) and analyze the impact of parametric driving on the full phase diagram of the dissipative Dicke model, see Fig.~\ref{fig:diagram}. 

The driven and dissipative nature of the system is  described by a Liouvillian equation for the density matrix $\rho_{\rm sys}$ of the system
\begin{eqnarray}\label{liouvillian}
\frac{d \rho_{\rm sys}}{dt} &=& - \frac{i}{\hbar} [H(t), \rho_{\rm sys}]  + \kappa[ 2 a \rho_{\rm sys} a^\dagger - \{ a^\dagger a , \rho_{\rm sys}\}]  \nonumber \\ &&+ 
\frac{\eta}{N}\sum_{i,j=1}^N  [ 2 s^i_- \rho_{\rm sys} s_+^j -\{s^i_+ s^j_-, \rho_{\rm sys}\}]\,,
\end{eqnarray}

where $s_{\pm}^j=s_{x}^j+is_{y}^j$ are ladder operators. The first term on the r.h.s. describes the standard Hamiltonian evolution and the last two terms
represent the Markovian dissipation for both cavity and a global dissipation for the atoms in Lindblad form with rates $\kappa$ and $\eta$, respectively~\cite{Miyashita2014}. Note that this approach is valid in the Born-Markov limit of weak dissipation. 
 
 \begin{table}
\begin{tabular}{|c|c|c|c|c|c|c|}
 \hline
            & Stability $A_1$    & Stability $A_2$ & $\left|\bar{\alpha}\right|$    &   $\left|\bar{x}\right|$  &  $\left|\bar{y}\right|$   & $\left|\bar{z}\right|$  \\
  \hline
  NP  & yes & yes & 0 & 0 & 0 & 1/2 \\
	\hline
  SP  & no & yes &  $\neq 0$ & $\neq 0$ & $\neq 0$ & $\neq 0$ \\
	\hline
  D-NP  & no ($\star$) & no ($\star$) &  $0$ & $0$ & $0$ & $< 1/2$ \\
  \hline
\end{tabular}
\caption{Summary of normal mode [cf.~Eq.~\eqref{cla-eom}] and mean field [cf.~Eqs.~\eqref{amean}-\eqref{ymean}] analyses. In Fig.~\ref{fig:diagram}(a), we observe three main regions, dubbed normal phase (NP), super-radiant phase (SP), and dynamical normal phase (D-NP). Each region manifests a different behavior summarized here, i.e.~, which normal mode is stable in each region~\cite{supmat}, and what are the values of the different order parameters. In the D-NP region, we denote by ``no ($\star$)'' that at least one normal mode is unstable.
} \label{table1}
\end{table}
 
In the absence of paramteric driving, $\epsilon=0$, the NP is well described by considering the collection of two-level atoms as constituting a giant spin ${\bf S}= \sum_i {\bf s}^i$ aligned along the $z$ axis~\cite{emary2003}. The deviations of this giant spin away from this quantization axis can be characterized by the standard Holstein Primakoff representation for the spin operators, $S_z =  b^\dagger b -{\frac N2} , S_-= \sqrt{N - b^\dagger b} b$, and $S_+= b^\dagger \sqrt{N - b^\dagger b} $, where $b, b^\dagger$ are standard bosonic operators~\cite{emary2003}. This approach can be extended to address the stability of the NP in the presence of parametric driving, $\epsilon\neq 0$.
Since deviations of $\mathbf{S}$ from the $z$ axis are expected to be small, as $N\to \infty$ we can map the
Dicke Hamiltonian [cf.~Eq.~\eqref{dickeHam}] onto the problem of two harmonic oscillators whose coupling is parametrically driven,
\begin{equation}\label{ham-np}
H_{\rm NP}(t)= \hbar\omega_c a^\dagger a + \hbar\omega_a b^\dagger b + \hbar\lambda(t) (b + b^\dagger) (a + a^\dagger)\,.
\end{equation}
%All higher order terms  in the Hamiltonian  scale as $1/N$ and can be dropped in the large $N$ limit.
Focusing  on the case  $\omega_0\equiv\omega_a=\omega_c$,  Eq.~(\ref{ham-np}) can be diagonalized in terms of normal modes, 
\begin{align}
\label{ham-nm}
H_{\rm NP}(t)= \hbar\Omega_1(t) A_1^\dagger A_1 + \hbar\Omega_2(t) A_2^\dagger A_2\,,
\end{align}
where for $m=1,2$, $\Omega_m^2(t)=\omega_0^2 - (-1)^{m} 2 \lambda(t) \omega_0$ are time-dependent  normal mode frequencies, and $A_m =\frac1{2\sqrt{2}} [ \mathcal{S}^+_{m} (a - (-1)^{m} b)+ \mathcal{S}^-_{m}(a^\dagger - (-1)^{m}b^\dagger )]$ are the corresponding normal mode operators with coefficients $\mathcal{S}_{m}^{\pm}= \frac{ \Omega_m \pm \omega_0}{2\sqrt{\omega_0 \Omega_m}}$. For computational simplicity, we also assume $\kappa=\eta\equiv\gamma$\cite{galve2010}.

Each normal mode is a quantum Mathieu parametric oscillator, i.e.~, its fundamental harmonic frequency varies sinusoidally in time~\cite{mclachlan1964,nayfehnonlinear}. The stability of each quantum parametric oscillator can be deduced from its displacement
${\rm Tr} \left\{\rho(t) (A_m + A_m^\dagger)\right\}$. It results in a complex stability diagram  comprising  ``Arnold tongues''  which delineate regions where the displacement, though parametrically amplified, remains bounded (stable), and those where the displacement grows exponentially with time (unstable)~\cite{zerbe1994,zerbe1995,supmat}. Incidentally, the resulting stability diagram for the quantum oscillator is the same as that of the classical damped Mathieu oscillator obeying the classical equations of motion for the displacement\cite{zerbe1994,zerbe1995,supmat},
\begin{eqnarray}\label{cla-eom}
{\ddot x}_m + \gamma \dot{x}_m + \Omega_{m}^2(t) x_m &=&0 \,,
\end{eqnarray}

The combination of the stability diagrams of the two normal modes yields the stability of the NP, i.e.~, the NP is stable
only if both dissipative normal modes are stable, see shaded area in Fig.~\ref{fig:diagram}.  
In the absence of dissipation, $\kappa=\eta=\gamma=0$, each normal mode $A_m$ is unstable in the limit of infinitesimal parametric driving $\epsilon\rightarrow 0$ at resonant frequencies $f_{m,n} = \sqrt{(\lambda_0)^2  + n^2 \Omega^2} + (-1)^m (\lambda_0)$ where $n= 1,2,3...$~\cite{mclachlan1964,nayfehnonlinear}. We find that the lower boundary of the NP is dictated by the lowest Arnold tongue of $A_2$, i.e.~, where the time-independent part of $\Omega_2(t)$ becomes negative. The remaining stability boundaries of NP are determined by frequencies where either $A_m$ becomes unstable.
The impact of disspation on the stability of NP can be understood from the physics of Mathieu oscillators where dissipation results in a modification of the stability criterion for the parametric oscillator. In particular, weak Markovian dissipation leads to a pronounced stabilization of the NP in the vicinity of these resonant frequencies for small driving $\epsilon\ll \lambda_0$, and barely affects the stability at higher drive amplitudes~\cite{zerbe1994,zerbe1995}. Indeed in Fig.~\ref{fig:diagram}, we see substantial stabilization of NP at the resonant frequency $f_{2,1}$. Such stabilization of the NP in the many-body context of the Dicke model is a manifestation  of the explicitly dissipation dependent non-equilibrium  asymptotic state.

From Fig.~\ref{fig:diagram}, we see that the NP occupies only a small part of the phase diagram when the system is parametrically driven. However, what lies beyond these stable NP regions cannot be accessed by the current approach and requires another method, such as mean field theory, which is well justified for the Dicke model in the limit of $N\gg 1$. This method was successfully used for studying the SP which has broken $Z_2$ symmetry, for the non-driven case $\epsilon = 0$~\cite{hepp1973superradiant,emary2003,Dimer2007}. The mean field ansatz that we use states that the total density matrix in the steady state is a product state of the individual density matrices, $\rho_{\rm sys} = \rho_c \otimes \prod_{i=1}^N \otimes \rho_i$, where $\rho_c$ and $\rho_i$ are density matrices of the cavity and the $i^{\rm th}$ atom, respectively~\cite{Miyashita2014}. Furthermore, since all atoms are identical, we assume all $\rho_i$ to be equivalent. Substituting this ansatz into Eq.~(\ref{liouvillian}), we obtain a set of coupled non-linear equations for the mean-field order parameters of the parametrically driven and dissipative Dicke model
\begin{align}
\dot \alpha=& -i  \omega_c \alpha - 2i \lambda(t)  x - \kappa \alpha \,,\label{amean}\\ 
{\dot x}=& - \omega_a y - \eta x \,,\label{xmean}\\
{\dot y}=&   \omega_a x - 2 \lambda(t) [\alpha + \alpha^*] z - \eta y\,,\label{ymean}
\end{align}
where we have defined the order parameters $\alpha=\langle a \rangle/\sqrt{N}$, $x=\langle \sum_i s^x_i \rangle/N$, $y=\langle \sum_i s^y_i \rangle/N$, $z=\langle \sum_i s^z_i \rangle/N$, and have assumed $z= \sqrt{(1/2)^2 - \vert x \vert^2 -\vert y \vert^2}$ since the mean field equations satisfy the constraint $x^2 + y^2 + z^2=1/4$.

Equations \eqref{amean}-\eqref{ymean} form a set of coupled non-autonomous differential equations. Using numerical stiff ordinary differential equations
solvers, we integrate this set of equations to a long time limit, where we obtain a convergent behavior. We find that, generically, the steady-state mean-field solutions show oscillatory behavior around a zero or non-zero mean value~\cite{supmat}. In Figs.~\ref{fig:diagram}(a), we plot the absolute time-averaged order parameters, $|\bar{\alpha}|,|\bar{x}|,|\bar{y}|$ and $|\bar{z}|$, averaged over a sufficiently long time window in the steady-state\cite{supmat}. Note, that the absolute value is taken for presentation reasons only, we always have $\bar{\alpha},\bar{y},\bar{z}>0$ and $\bar{x}<0$. Based on these mean values, 
we find that the SP, characterized by $\bar{\alpha}\neq 0$, extends from its zero drive region of $\omega_0 \leq 2 \lambda_0$ onto a large regime spanning both small and large drive amplitudes. At the critical frequency $\omega_{\rm crit}(\omega_0,\epsilon)$, we observe a transition to
a region with $\bar{\alpha}=\bar{x}=\bar{y}= 0$. Contrasting these results with those obtained through the study of normal modes, we see that for $\epsilon  \lesssim 0.3 \Omega$, the NP lies above the line defined by $\omega_{\rm crit}$ with $\bar{z}=1/2$. The NP$\leftrightarrow$SP transition in this regime is
thus an extension of the usual continuous Dicke transition at zero drive to finite  parametric driving. Note that the details of the transition may still differ from the standard Dicke transition as the parametrically-driven NP accommodates a large number of photons in the cavity.

Interestingly, at $\epsilon\sim 0.3 \Omega$, we see a sudden change in the curvature of $\omega_{\rm crit}$. This exactly signals the point where the NP ends and a novel dynamical phase, dubbed dynamical-NP (D-NP), starts. As opposed to the NP,  though $\bar\alpha=0$, this phase has  oscillatory $\alpha(t)$  and does not have its spin aligned along the $z$-axis, i.e.~, $\bar{z}<1/2$. Additionally, this region corresponds exactly to the parametrically unstable Arnold tongues of the aforementioned normal modes. As a result, the point $(\epsilon \sim 0.3 \Omega, \omega_{\rm crit})$, appears to be a multicritical point where the three phases: NP, SP, and the new D-NP intersect. The principal features of the three phases are summarized in Table \ref{table1}.

\begin{figure}[htbp]                   
\centering
\includegraphics[width=50mm]{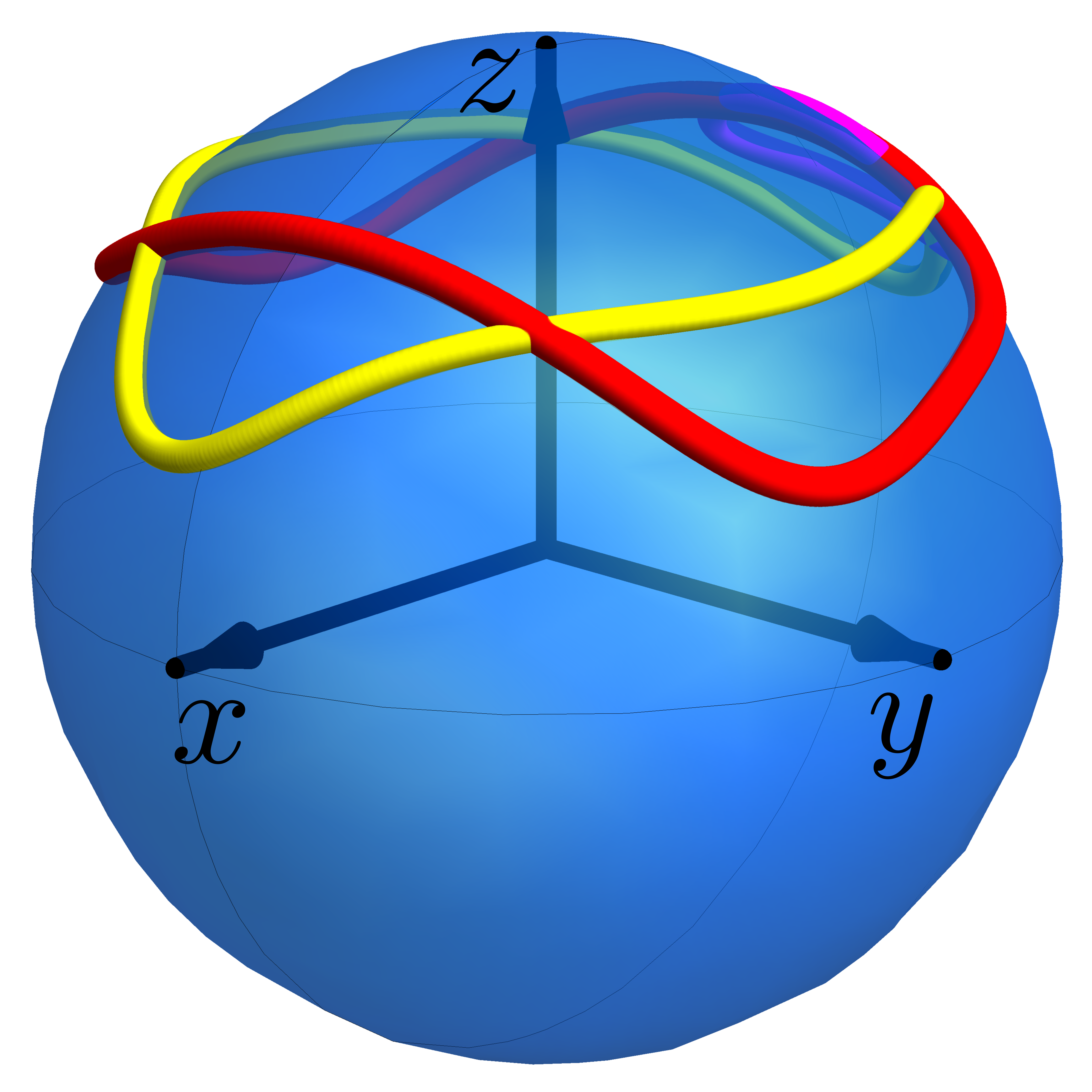}
\caption{ \label{fig:bloch} Trajectories on the Bloch sphere of the atoms order parameters as a function of time in steady-state, $f[x(t), y(t), z(t),t]$. The trajectories are for $\epsilon=0.5\Omega$, $\lambda_0=0.4\Omega$, and $\kappa=\eta=0.1\Omega$. The trajectory that does not encircle the $z$-axis (magenta) is in SP with $\omega_0=0.6\Omega$. The trajectories that encircle the $z$-axis (red and yellow) are for $\omega_0 = 0.65\Omega$ and $\omega_0 = 1.5\Omega$, respectively.}
\end{figure}

%In the absence of driving,i.e., $\epsilon=0$,  the order parameters converge to their steady state values with $\dot \alpha= {\dot x} ={\dot y}=0$. In the steady state, 
%it is straightforward to obtain the order parameters\ cite{bastidas2012, dimer} and  
%\begin{eqnarray}
%\Re[\alpha] & = &\frac{1}{16\lambda_0^2( \omega^2 + \eta^2)}[\frac{16 \lambda_0^4 \omega^4}{( \omega^2 + \kappa^2)^2} - {( \omega^2 + \eta^2)^2}] \\
%x^2 & = &\frac{1}{64\lambda_0^4 \omega^2 ( \omega^2 + \eta^2)}[16 \lambda_0^4 \omega^4 - ( \omega^2 + \kappa^2)^2 ( \omega^2 + \eta^2)^2] \\
%y&=& -\eta \frac{x}{\omega} 
%\end{eqnarray}
%When $\epsilon\neq 0$, the full mean field equations have to be solved to obtain the steady state which is oscillatory with a periodicity related to the
%underlying period $T= 2\pi/\Omega$. 
%The normal phase behaviour can also be obtained by  rewriting the  normal mode Hamiltonian as
%\begin{equation} \label{aham-1}
%H_i = [\omega_{oi} + \frac{\chi^2(t)}{2 \omega_0}] a^\dagger a + \frac{\phi^2_i(t)}{4 \omega_0} (a^2 + a^{\dagger 2})
%\equiv \sqrt{\omega^2_{oi} + \chi^2_i(t)} {\bar a}^\dagger {\bar a}
%\end{equation}
%where 
%$\chi^2_{1,2}(t)= \pm q  \cos(2\Omega t + \phi)$.  The equations of motion of the operators can be  solved to obtain the
To better understand the nature of the D-NP, as well as the effects of the drive $\epsilon\neq 0$ on the NP and SP, we analyze the oscillatory behaviour around the steady-state mean-field solutions of Eqs.~\eqref{amean}-\eqref{ymean}. Typically, we find that each phase has a different oscillatory behavior: (i) in NP, the order parameters converge to zero and do not oscillate, (ii) in SP, the oscillations are small but mildly grow with $\epsilon$, and (iii) in D-NP, the order parameters oscillate strongly around zero \cite{supmat}. We, then, Fast Fourier Transform (FFT) the steady-state solutions for each $\omega_0$ and $\epsilon$. In both SP and D-NP regimes, the oscillations have a complex beat structure that corresponds to a comb of peaked frequencies at $\omega\neq 0$, as well as $\omega\neq \Omega, \omega_0$ \cite{supmat}. Note that the frequencies that appear can be understood from Floquet analysis~\cite{morse1953methods}. In Figs.~\ref{fig:diagram}(b), we plot the amplitude of the largest $\omega\neq 0$ peak in the FFT landscape in order to quantify the overall extent of the oscillation. We find that, SP has weak oscillations in all of the order parameters, whereas the D-NP is strongly oscillatory in the $x-y$ plane. These aspects are better highlighted in  Fig.~\ref{fig:bloch}, by plotting the steady-state time-dependent trajectory of the total spin $\langle \sum_i {\bf s}^i |(t)\rangle$ on the Bloch sphere for different parameter values. It appears that a distinguishing criterion between the SP and D-NP is whether the trajectory encircles the $z$-axis. Our results seem to indicate that the most plausible candidate for the D-NP is a "normal phase" in a dynamical rotating frame. 

Combining the results from the normal modes and mean field analyses, we see that periodic modulation  of the atom-light coupling  results in a rich phase diagram, characterized by a multitude of dynamical phase boundaries   between the NP, SP and the intriguing new phase, D-NP. All   three phases  meet at  a multicritical point.   In the D-NP,  the cavity  periodically emits pulses of photons  with opposing phases, which should be detectable experimentally.
 The  NP $\rightarrow$ SP  boundary is principally dictated by where the normal mode $A_2$ becomes unstable, whereas  the NP $\rightarrow$ D-NP  boundary is fixed by the  instability of any of the modes $A_m$  \cite{supmat}.  Within our mean field
approach, we find the transitions  SP $\rightarrow$ NP, D-NP to be continuous, though the latter is rather sharp.  However, the nature of the NP $\rightarrow$ D-NP  transition 
 cannot be studied within our approach. The topography  of the phase diagram is expected to  vary with the choice of $\lambda_0$ and the strength of  dissipation. Consequently, though the three phases would exist, the 
 highly sensitive multicritical point may  disappear.
 
Remarkably, we see that dissipation  leads  to  a sizable stabilization of the NP.  Due to the parametric nature of the normal modes, the NP in the driven
 case can manifest a dissipation-assisted generation of substantial entanglement/squeezing between the atoms of the condensate and the cavity~\cite{galve2010}.  The physical signature of such entanglement as well as the impact of parametric driving   and dissipation on the critical exponents defining the different phase transitions
%diverging fluctuations as well as basic  excitations in all the phases  
merit in-depth studies. It would also be interesting to extend the present work to other parameter regimes like  $\omega_a << \omega_c$ realized in current experimental setups~\cite{brennecke2013}. 

Our work shows that parametric driving is a powerful tool in the quest for new physics, which exists exclusively in  the realm  far from equilibrium.  The richness of the physics seen in the  simple Dicke model presages intriguing phenomena in time-dependent systems, which requires the development of new theoretical methodologies. This frontier is potentially best explored using experimental light-matter systems. 

We would like to thank  T. Donner, R. Mottl, R. Landig, T. Esslinger, L. Papariello, and E. van Nieuwenburg for useful discussions. We acknowledge financial support from the Swiss National Science Foundation (SNSF).

%\bibliography{prmDrive}

\begin{thebibliography}{28}%
\makeatletter
\providecommand \@ifxundefined [1]{%
 \@ifx{#1\undefined}
}%
\providecommand \@ifnum [1]{%
 \ifnum #1\expandafter \@firstoftwo
 \else \expandafter \@secondoftwo
 \fi
}%
\providecommand \@ifx [1]{%
 \ifx #1\expandafter \@firstoftwo
 \else \expandafter \@secondoftwo
 \fi
}%
\providecommand \natexlab [1]{#1}%
\providecommand \enquote  [1]{``#1''}%
\providecommand \bibnamefont  [1]{#1}%
\providecommand \bibfnamefont [1]{#1}%
\providecommand \citenamefont [1]{#1}%
\providecommand \href@noop [0]{\@secondoftwo}%
\providecommand \href [0]{\begingroup \@sanitize@url \@href}%
\providecommand \@href[1]{\@@startlink{#1}\@@href}%
\providecommand \@@href[1]{\endgroup#1\@@endlink}%
\providecommand \@sanitize@url [0]{\catcode `\\12\catcode `\$12\catcode
  `\&12\catcode `\#12\catcode `\^12\catcode `\_12\catcode `\%12\relax}%
\providecommand \@@startlink[1]{}%
\providecommand \@@endlink[0]{}%
\providecommand \url  [0]{\begingroup\@sanitize@url \@url }%
\providecommand \@url [1]{\endgroup\@href {#1}{\urlprefix }}%
\providecommand \urlprefix  [0]{URL }%
\providecommand \Eprint [0]{\href }%
\providecommand \doibase [0]{http://dx.doi.org/}%
\providecommand \selectlanguage [0]{\@gobble}%
\providecommand \bibinfo  [0]{\@secondoftwo}%
\providecommand \bibfield  [0]{\@secondoftwo}%
\providecommand \translation [1]{[#1]}%
\providecommand \BibitemOpen [0]{}%
\providecommand \bibitemStop [0]{}%
\providecommand \bibitemNoStop [0]{.\EOS\space}%
\providecommand \EOS [0]{\spacefactor3000\relax}%
\providecommand \BibitemShut  [1]{\csname bibitem#1\endcsname}%
\let\auto@bib@innerbib\@empty
%</preamble>
\bibitem [{\citenamefont {Carusotto}\ and\ \citenamefont
  {Ciuti}(2013)}]{Carusotto2013}%
  \BibitemOpen
  \bibfield  {author} {\bibinfo {author} {\bibfnamefont {I.}~\bibnamefont
  {Carusotto}}\ and\ \bibinfo {author} {\bibfnamefont {C.}~\bibnamefont
  {Ciuti}},\ }\href@noop {} {\bibfield  {journal} {\bibinfo  {journal} {Rev.
  Mod. Phys.}\ }\textbf {\bibinfo {volume} {85}},\ \bibinfo {pages} {299}
  (\bibinfo {year} {2013})}\BibitemShut {NoStop}%
\bibitem [{\citenamefont {Bloch}\ \emph {et~al.}(2008)\citenamefont {Bloch},
  \citenamefont {Dalibard},\ and\ \citenamefont {Zwerger}}]{Bloch2008}%
  \BibitemOpen
  \bibfield  {author} {\bibinfo {author} {\bibfnamefont {I.}~\bibnamefont
  {Bloch}}, \bibinfo {author} {\bibfnamefont {J.}~\bibnamefont {Dalibard}}, \
  and\ \bibinfo {author} {\bibfnamefont {W.}~\bibnamefont {Zwerger}},\
  }\href@noop {} {\bibfield  {journal} {\bibinfo  {journal} {Rev. Mod. Phys.}\
  }\textbf {\bibinfo {volume} {80}},\ \bibinfo {pages} {885} (\bibinfo {year}
  {2008})}\BibitemShut {NoStop}%
\bibitem [{\citenamefont {Kavokin}\ \emph {et~al.}(2007)\citenamefont
  {Kavokin}, \citenamefont {Baumberg}, \citenamefont {Malpuech},\ and\
  \citenamefont {Laussy}}]{kavokin2007microcavities}%
  \BibitemOpen
  \bibfield  {author} {\bibinfo {author} {\bibfnamefont {A.}~\bibnamefont
  {Kavokin}}, \bibinfo {author} {\bibfnamefont {J.~J.}\ \bibnamefont
  {Baumberg}}, \bibinfo {author} {\bibfnamefont {G.}~\bibnamefont {Malpuech}},
  \ and\ \bibinfo {author} {\bibfnamefont {F.~P.}\ \bibnamefont {Laussy}},\
  }\href@noop {} {\emph {\bibinfo {title} {Microcavities}}}\ (\bibinfo
  {publisher} {Oxford University Press},\ \bibinfo {year} {2007})\BibitemShut
  {NoStop}%
\bibitem [{\citenamefont {Wallraff}\ \emph {et~al.}(2004)\citenamefont
  {Wallraff}, \citenamefont {Schuster}, \citenamefont {Blais}, \citenamefont
  {Frunzio}, \citenamefont {Huang}, \citenamefont {Majer}, \citenamefont
  {Kumar}, \citenamefont {Girvin},\ and\ \citenamefont
  {Schoelkopf}}]{wallraff2004strong}%
  \BibitemOpen
  \bibfield  {author} {\bibinfo {author} {\bibfnamefont {A.}~\bibnamefont
  {Wallraff}}, \bibinfo {author} {\bibfnamefont {D.~I.}\ \bibnamefont
  {Schuster}}, \bibinfo {author} {\bibfnamefont {A.}~\bibnamefont {Blais}},
  \bibinfo {author} {\bibfnamefont {L.}~\bibnamefont {Frunzio}}, \bibinfo
  {author} {\bibfnamefont {R.-S.}\ \bibnamefont {Huang}}, \bibinfo {author}
  {\bibfnamefont {J.}~\bibnamefont {Majer}}, \bibinfo {author} {\bibfnamefont
  {S.}~\bibnamefont {Kumar}}, \bibinfo {author} {\bibfnamefont {S.~M.}\
  \bibnamefont {Girvin}}, \ and\ \bibinfo {author} {\bibfnamefont {R.~J.}\
  \bibnamefont {Schoelkopf}},\ }\href@noop {} {\bibfield  {journal} {\bibinfo
  {journal} {Nature}\ }\textbf {\bibinfo {volume} {431}},\ \bibinfo {pages}
  {162} (\bibinfo {year} {2004})}\BibitemShut {NoStop}%
\bibitem [{\citenamefont {Majer}\ \emph {et~al.}(2007)\citenamefont {Majer},
  \citenamefont {Chow}, \citenamefont {Gambetta}, \citenamefont {Koch},
  \citenamefont {Johnson}, \citenamefont {Schreier}, \citenamefont {Frunzio},
  \citenamefont {Schuster}, \citenamefont {Houck}, \citenamefont {Wallraff}
  \emph {et~al.}}]{majer2007coupling}%
  \BibitemOpen
  \bibfield  {author} {\bibinfo {author} {\bibfnamefont {J.}~\bibnamefont
  {Majer}}, \bibinfo {author} {\bibfnamefont {J.}~\bibnamefont {Chow}},
  \bibinfo {author} {\bibfnamefont {J.}~\bibnamefont {Gambetta}}, \bibinfo
  {author} {\bibfnamefont {J.}~\bibnamefont {Koch}}, \bibinfo {author}
  {\bibfnamefont {B.}~\bibnamefont {Johnson}}, \bibinfo {author} {\bibfnamefont
  {J.}~\bibnamefont {Schreier}}, \bibinfo {author} {\bibfnamefont
  {L.}~\bibnamefont {Frunzio}}, \bibinfo {author} {\bibfnamefont
  {D.}~\bibnamefont {Schuster}}, \bibinfo {author} {\bibfnamefont
  {A.}~\bibnamefont {Houck}}, \bibinfo {author} {\bibfnamefont
  {A.}~\bibnamefont {Wallraff}},  \emph {et~al.},\ }\href@noop {} {\bibfield
  {journal} {\bibinfo  {journal} {Nature}\ }\textbf {\bibinfo {volume} {449}},\
  \bibinfo {pages} {443} (\bibinfo {year} {2007})}\BibitemShut {NoStop}%
\bibitem [{\citenamefont {Amo}\ \emph {et~al.}(2009)\citenamefont {Amo},
  \citenamefont {Lefr{\`e}re}, \citenamefont {Pigeon}, \citenamefont {Adrados},
  \citenamefont {Ciuti}, \citenamefont {Carusotto}, \citenamefont {Houdr{\'e}},
  \citenamefont {Giacobino},\ and\ \citenamefont
  {Bramati}}]{amo2009superfluidity}%
  \BibitemOpen
  \bibfield  {author} {\bibinfo {author} {\bibfnamefont {A.}~\bibnamefont
  {Amo}}, \bibinfo {author} {\bibfnamefont {J.}~\bibnamefont {Lefr{\`e}re}},
  \bibinfo {author} {\bibfnamefont {S.}~\bibnamefont {Pigeon}}, \bibinfo
  {author} {\bibfnamefont {C.}~\bibnamefont {Adrados}}, \bibinfo {author}
  {\bibfnamefont {C.}~\bibnamefont {Ciuti}}, \bibinfo {author} {\bibfnamefont
  {I.}~\bibnamefont {Carusotto}}, \bibinfo {author} {\bibfnamefont
  {R.}~\bibnamefont {Houdr{\'e}}}, \bibinfo {author} {\bibfnamefont
  {E.}~\bibnamefont {Giacobino}}, \ and\ \bibinfo {author} {\bibfnamefont
  {A.}~\bibnamefont {Bramati}},\ }\href@noop {} {\bibfield  {journal} {\bibinfo
   {journal} {Nature Physics}\ }\textbf {\bibinfo {volume} {5}},\ \bibinfo
  {pages} {805} (\bibinfo {year} {2009})}\BibitemShut {NoStop}%
\bibitem [{\citenamefont {Brennecke}\ \emph {et~al.}(2013)\citenamefont
  {Brennecke}, \citenamefont {Mottl}, \citenamefont {Baumann}, \citenamefont
  {Landig}, \citenamefont {Donner},\ and\ \citenamefont
  {Esslinger}}]{brennecke2013}%
  \BibitemOpen
  \bibfield  {author} {\bibinfo {author} {\bibfnamefont {F.}~\bibnamefont
  {Brennecke}}, \bibinfo {author} {\bibfnamefont {R.}~\bibnamefont {Mottl}},
  \bibinfo {author} {\bibfnamefont {K.}~\bibnamefont {Baumann}}, \bibinfo
  {author} {\bibfnamefont {R.}~\bibnamefont {Landig}}, \bibinfo {author}
  {\bibfnamefont {T.}~\bibnamefont {Donner}}, \ and\ \bibinfo {author}
  {\bibfnamefont {T.}~\bibnamefont {Esslinger}},\ }\href@noop {} {\bibfield
  {journal} {\bibinfo  {journal} {Proceedings of the National Academy of
  Sciences}\ }\textbf {\bibinfo {volume} {110}},\ \bibinfo {pages} {11763}
  (\bibinfo {year} {2013})}\BibitemShut {NoStop}%
\bibitem [{\citenamefont {Rechtsman}\ \emph {et~al.}(2013)\citenamefont
  {Rechtsman}, \citenamefont {Zeuner}, \citenamefont {Plotnik}, \citenamefont
  {Lumer}, \citenamefont {Podolsky}, \citenamefont {Dreisow}, \citenamefont
  {Nolte}, \citenamefont {Segev},\ and\ \citenamefont {Szameit}}]{Floquet}%
  \BibitemOpen
  \bibfield  {author} {\bibinfo {author} {\bibfnamefont {M.~C.}\ \bibnamefont
  {Rechtsman}}, \bibinfo {author} {\bibfnamefont {J.~M.}\ \bibnamefont
  {Zeuner}}, \bibinfo {author} {\bibfnamefont {Y.}~\bibnamefont {Plotnik}},
  \bibinfo {author} {\bibfnamefont {Y.}~\bibnamefont {Lumer}}, \bibinfo
  {author} {\bibfnamefont {D.}~\bibnamefont {Podolsky}}, \bibinfo {author}
  {\bibfnamefont {F.}~\bibnamefont {Dreisow}}, \bibinfo {author} {\bibfnamefont
  {S.}~\bibnamefont {Nolte}}, \bibinfo {author} {\bibfnamefont
  {M.}~\bibnamefont {Segev}}, \ and\ \bibinfo {author} {\bibfnamefont
  {A.}~\bibnamefont {Szameit}},\ }\href@noop {} {\bibfield  {journal} {\bibinfo
   {journal} {Nature}\ }\textbf {\bibinfo {volume} {496}},\ \bibinfo {pages}
  {196} (\bibinfo {year} {2013})}\BibitemShut {NoStop}%
\bibitem [{\citenamefont {Orzel}\ \emph {et~al.}(2001)\citenamefont {Orzel},
  \citenamefont {Tuchman}, \citenamefont {Fenselau}, \citenamefont {Yasuda},\
  and\ \citenamefont {Kasevich}}]{orzel2001squeezed}%
  \BibitemOpen
  \bibfield  {author} {\bibinfo {author} {\bibfnamefont {C.}~\bibnamefont
  {Orzel}}, \bibinfo {author} {\bibfnamefont {A.}~\bibnamefont {Tuchman}},
  \bibinfo {author} {\bibfnamefont {M.}~\bibnamefont {Fenselau}}, \bibinfo
  {author} {\bibfnamefont {M.}~\bibnamefont {Yasuda}}, \ and\ \bibinfo {author}
  {\bibfnamefont {M.}~\bibnamefont {Kasevich}},\ }\href@noop {} {\bibfield
  {journal} {\bibinfo  {journal} {Science}\ }\textbf {\bibinfo {volume}
  {291}},\ \bibinfo {pages} {2386} (\bibinfo {year} {2001})}\BibitemShut
  {NoStop}%
\bibitem [{\citenamefont {Rugar}\ \emph {et~al.}(2004)\citenamefont {Rugar},
  \citenamefont {Budakian}, \citenamefont {Mamin},\ and\ \citenamefont
  {Chui}}]{rugar2004single}%
  \BibitemOpen
  \bibfield  {author} {\bibinfo {author} {\bibfnamefont {D.}~\bibnamefont
  {Rugar}}, \bibinfo {author} {\bibfnamefont {R.}~\bibnamefont {Budakian}},
  \bibinfo {author} {\bibfnamefont {H.}~\bibnamefont {Mamin}}, \ and\ \bibinfo
  {author} {\bibfnamefont {B.}~\bibnamefont {Chui}},\ }\href@noop {} {\bibfield
   {journal} {\bibinfo  {journal} {Nature}\ }\textbf {\bibinfo {volume}
  {430}},\ \bibinfo {pages} {329} (\bibinfo {year} {2004})}\BibitemShut
  {NoStop}%
\bibitem [{\citenamefont {G{\"u}nter}\ \emph {et~al.}(2009)\citenamefont
  {G{\"u}nter}, \citenamefont {Anappara}, \citenamefont {Hees}, \citenamefont
  {Sell}, \citenamefont {Biasiol}, \citenamefont {Sorba}, \citenamefont
  {De~Liberato}, \citenamefont {Ciuti}, \citenamefont {Tredicucci},
  \citenamefont {Leitenstorfer} \emph {et~al.}}]{QED}%
  \BibitemOpen
  \bibfield  {author} {\bibinfo {author} {\bibfnamefont {G.}~\bibnamefont
  {G{\"u}nter}}, \bibinfo {author} {\bibfnamefont {A.}~\bibnamefont
  {Anappara}}, \bibinfo {author} {\bibfnamefont {J.}~\bibnamefont {Hees}},
  \bibinfo {author} {\bibfnamefont {A.}~\bibnamefont {Sell}}, \bibinfo {author}
  {\bibfnamefont {G.}~\bibnamefont {Biasiol}}, \bibinfo {author} {\bibfnamefont
  {L.}~\bibnamefont {Sorba}}, \bibinfo {author} {\bibfnamefont
  {S.}~\bibnamefont {De~Liberato}}, \bibinfo {author} {\bibfnamefont
  {C.}~\bibnamefont {Ciuti}}, \bibinfo {author} {\bibfnamefont
  {A.}~\bibnamefont {Tredicucci}}, \bibinfo {author} {\bibfnamefont
  {A.}~\bibnamefont {Leitenstorfer}},  \emph {et~al.},\ }\href@noop {}
  {\bibfield  {journal} {\bibinfo  {journal} {Nature}\ }\textbf {\bibinfo
  {volume} {458}},\ \bibinfo {pages} {178} (\bibinfo {year}
  {2009})}\BibitemShut {NoStop}%
\bibitem [{\citenamefont {Dimer}\ \emph {et~al.}(2007)\citenamefont {Dimer},
  \citenamefont {Estienne}, \citenamefont {Parkins},\ and\ \citenamefont
  {Carmichael}}]{Dimer2007}%
  \BibitemOpen
  \bibfield  {author} {\bibinfo {author} {\bibfnamefont {F.}~\bibnamefont
  {Dimer}}, \bibinfo {author} {\bibfnamefont {B.}~\bibnamefont {Estienne}},
  \bibinfo {author} {\bibfnamefont {A.~S.}\ \bibnamefont {Parkins}}, \ and\
  \bibinfo {author} {\bibfnamefont {H.~J.}\ \bibnamefont {Carmichael}},\ }\href
  {\doibase 10.1103/PhysRevA.75.013804} {\bibfield  {journal} {\bibinfo
  {journal} {Phys. Rev. A}\ }\textbf {\bibinfo {volume} {75}},\ \bibinfo
  {pages} {013804} (\bibinfo {year} {2007})}\BibitemShut {NoStop}%
\bibitem [{\citenamefont {Klinder}\ \emph {et~al.}(2014)\citenamefont
  {Klinder}, \citenamefont {Ke{\ss}ler}, \citenamefont {Wolke}, \citenamefont
  {Mathey},\ and\ \citenamefont {Hemmerich}}]{hemmerich2014}%
  \BibitemOpen
  \bibfield  {author} {\bibinfo {author} {\bibfnamefont {J.}~\bibnamefont
  {Klinder}}, \bibinfo {author} {\bibfnamefont {H.}~\bibnamefont {Ke{\ss}ler}},
  \bibinfo {author} {\bibfnamefont {M.}~\bibnamefont {Wolke}}, \bibinfo
  {author} {\bibfnamefont {L.}~\bibnamefont {Mathey}}, \ and\ \bibinfo {author}
  {\bibfnamefont {A.}~\bibnamefont {Hemmerich}},\ }\href@noop {} {\bibfield
  {journal} {\bibinfo  {journal} {arXiv preprint arXiv:1409.1945}\ } (\bibinfo
  {year} {2014})}\BibitemShut {NoStop}%
\bibitem [{\citenamefont {Nagy}\ \emph {et~al.}(2011)\citenamefont {Nagy},
  \citenamefont {Szirmai},\ and\ \citenamefont {Domokos}}]{Nagy2011}%
  \BibitemOpen
  \bibfield  {author} {\bibinfo {author} {\bibfnamefont {D.}~\bibnamefont
  {Nagy}}, \bibinfo {author} {\bibfnamefont {G.}~\bibnamefont {Szirmai}}, \
  and\ \bibinfo {author} {\bibfnamefont {P.}~\bibnamefont {Domokos}},\ }\href
  {\doibase 10.1103/PhysRevA.84.043637} {\bibfield  {journal} {\bibinfo
  {journal} {Phys. Rev. A}\ }\textbf {\bibinfo {volume} {84}},\ \bibinfo
  {pages} {043637} (\bibinfo {year} {2011})}\BibitemShut {NoStop}%
\bibitem [{sup()}]{supmat}%
  \BibitemOpen
  \href@noop {} {}\bibinfo {note} {See Supplemental Material for additional
  details.}\BibitemShut {Stop}%
\bibitem [{eps()}]{epsilon}%
  \BibitemOpen
  \href@noop {} {}\bibinfo {note} {Though $\epsilon \le \lambda_0$ in current
  experimental setups, we explore the impact of larger drive amplitudes as
  well.}\BibitemShut {Stop}%
\bibitem [{\citenamefont {Bastidas}\ \emph {et~al.}(2012)\citenamefont
  {Bastidas}, \citenamefont {Emary}, \citenamefont {Regler},\ and\
  \citenamefont {Brandes}}]{bastidas2012}%
  \BibitemOpen
  \bibfield  {author} {\bibinfo {author} {\bibfnamefont {V.~M.}~\bibnamefont
  {Bastidas}}, \bibinfo {author} {\bibfnamefont {C.}~\bibnamefont {Emary}},
  \bibinfo {author} {\bibfnamefont {B.}~\bibnamefont {Regler}}, \ and\ \bibinfo
  {author} {\bibfnamefont {T.}~\bibnamefont {Brandes}},\ }\href@noop {}
  {\bibfield  {journal} {\bibinfo  {journal} {Phys. Rev. Lett.}\ }\textbf
  {\bibinfo {volume} {108}},\ \bibinfo {pages} {043003} (\bibinfo {year}
  {2012})}\BibitemShut {NoStop}%
\bibitem [{\citenamefont {Vacanti}\ \emph {et~al.}(2012)\citenamefont
  {Vacanti}, \citenamefont {Pugnetti}, \citenamefont {Didier}, \citenamefont
  {Paternostro}, \citenamefont {Palma}, \citenamefont {Fazio},\ and\
  \citenamefont {Vedral}}]{vacanti2012}%
  \BibitemOpen
  \bibfield  {author} {\bibinfo {author} {\bibfnamefont {G.}~\bibnamefont
  {Vacanti}}, \bibinfo {author} {\bibfnamefont {S.}~\bibnamefont {Pugnetti}},
  \bibinfo {author} {\bibfnamefont {N.}~\bibnamefont {Didier}}, \bibinfo
  {author} {\bibfnamefont {M.}~\bibnamefont {Paternostro}}, \bibinfo {author}
  {\bibfnamefont {G.~M.}~\bibnamefont {Palma}}, \bibinfo {author} {\bibfnamefont
  {R.}~\bibnamefont {Fazio}}, \ and\ \bibinfo {author} {\bibfnamefont
  {V.}~\bibnamefont {Vedral}},\ }\href@noop {} {\bibfield  {journal} {\bibinfo
  {journal} {Phys. Rev. Lett.}\ }\textbf {\bibinfo {volume} {108}},\ \bibinfo
  {pages} {093603} (\bibinfo {year} {2012})}\BibitemShut {NoStop}%
\bibitem [{\citenamefont {Shirai}\ \emph {et~al.}(2014)\citenamefont {Shirai},
  \citenamefont {Mori},\ and\ \citenamefont {Miyashita}}]{Miyashita2014}%
  \BibitemOpen
  \bibfield  {author} {\bibinfo {author} {\bibfnamefont {T.}~\bibnamefont
  {Shirai}}, \bibinfo {author} {\bibfnamefont {T.}~\bibnamefont {Mori}}, \ and\
  \bibinfo {author} {\bibfnamefont {S.}~\bibnamefont {Miyashita}},\ }\href@noop
  {} {\bibfield  {journal} {\bibinfo  {journal} {Journal of Physics B: Atomic,
  Molecular and Optical Physics}\ }\textbf {\bibinfo {volume} {47}},\ \bibinfo
  {pages} {025501} (\bibinfo {year} {2014})}\BibitemShut {NoStop}%
\bibitem [{\citenamefont {Emary}\ and\ \citenamefont
  {Brandes}(2003)}]{emary2003}%
  \BibitemOpen
  \bibfield  {author} {\bibinfo {author} {\bibfnamefont {C.}~\bibnamefont
  {Emary}}\ and\ \bibinfo {author} {\bibfnamefont {T.}~\bibnamefont
  {Brandes}},\ }\href@noop {} {\bibfield  {journal} {\bibinfo  {journal} {Phys.
  Rev. E}\ }\textbf {\bibinfo {volume} {67}},\ \bibinfo {pages} {066203}
  (\bibinfo {year} {2003})}\BibitemShut {NoStop}%
\bibitem [{\citenamefont {Galve}\ \emph {et~al.}(2010)\citenamefont {Galve},
  \citenamefont {Pach{\'o}n},\ and\ \citenamefont {Zueco}}]{galve2010}%
  \BibitemOpen
  \bibfield  {author} {\bibinfo {author} {\bibfnamefont {F.}~\bibnamefont
  {Galve}}, \bibinfo {author} {\bibfnamefont {L.~A.}\ \bibnamefont
  {Pach{\'o}n}}, \ and\ \bibinfo {author} {\bibfnamefont {D.}~\bibnamefont
  {Zueco}},\ }\href@noop {} {\bibfield  {journal} {\bibinfo  {journal} {Phys.
  Rev. Lett.}\ }\textbf {\bibinfo {volume} {105}},\ \bibinfo {pages} {180501}
  (\bibinfo {year} {2010})}\BibitemShut {NoStop}%
\bibitem [{\citenamefont {McLachlan}(1964)}]{mclachlan1964}%
  \BibitemOpen
  \bibfield  {author} {\bibinfo {author} {\bibfnamefont {N.~W.}\ \bibnamefont
  {McLachlan}},\ }\href@noop {} {\emph {\bibinfo {title} {Theory and
  application of Mathieu functions}}}\ (\bibinfo  {publisher} {Dover, New
  York},\ \bibinfo {year} {1964})\BibitemShut {NoStop}%
\bibitem [{\citenamefont {Nayfeh}\ and\ \citenamefont
  {Mook}(1979)}]{nayfehnonlinear}%
  \BibitemOpen
  \bibfield  {author} {\bibinfo {author} {\bibfnamefont {A.~H.}\ \bibnamefont
  {Nayfeh}}\ and\ \bibinfo {author} {\bibfnamefont {D.}~\bibnamefont {Mook}},\
  }\href@noop {} {\emph {\bibinfo {title} {Nonlinear oscillations}}}\ (\bibinfo
   {publisher} {Willey, New York},\ \bibinfo {year} {1979})\BibitemShut
  {NoStop}%
\bibitem [{\citenamefont {Zerbe}\ \emph {et~al.}(1994)\citenamefont {Zerbe},
  \citenamefont {Jung},\ and\ \citenamefont {H{\"a}nggi}}]{zerbe1994}%
  \BibitemOpen
  \bibfield  {author} {\bibinfo {author} {\bibfnamefont {C.}~\bibnamefont
  {Zerbe}}, \bibinfo {author} {\bibfnamefont {P.}~\bibnamefont {Jung}}, \ and\
  \bibinfo {author} {\bibfnamefont {P.}~\bibnamefont {H{\"a}nggi}},\
  }\href@noop {} {\bibfield  {journal} {\bibinfo  {journal} {Phys. Rev. E}\
  }\textbf {\bibinfo {volume} {49}},\ \bibinfo {pages} {3626} (\bibinfo {year}
  {1994})}\BibitemShut {NoStop}%
\bibitem [{\citenamefont {Zerbe}\ and\ \citenamefont
  {H{\"a}nggi}(1995)}]{zerbe1995}%
  \BibitemOpen
  \bibfield  {author} {\bibinfo {author} {\bibfnamefont {C.}~\bibnamefont
  {Zerbe}}\ and\ \bibinfo {author} {\bibfnamefont {P.}~\bibnamefont
  {H{\"a}nggi}},\ }\href@noop {} {\bibfield  {journal} {\bibinfo  {journal}
  {Phys. Rev. E}\ }\textbf {\bibinfo {volume} {52}},\ \bibinfo {pages} {1533}
  (\bibinfo {year} {1995})}\BibitemShut {NoStop}%
\bibitem [{\citenamefont {Hepp}\ and\ \citenamefont
  {Lieb}(1973)}]{hepp1973superradiant}%
  \BibitemOpen
  \bibfield  {author} {\bibinfo {author} {\bibfnamefont {K.}~\bibnamefont
  {Hepp}}\ and\ \bibinfo {author} {\bibfnamefont {E.~H.}\ \bibnamefont
  {Lieb}},\ }\href@noop {} {\bibfield  {journal} {\bibinfo  {journal} {Annals
  of Physics}\ }\textbf {\bibinfo {volume} {76}},\ \bibinfo {pages} {360}
  (\bibinfo {year} {1973})}\BibitemShut {NoStop}%
\bibitem [{\citenamefont {Morse}\ and\ \citenamefont
  {Feshbach}(1953)}]{morse1953methods}%
  \BibitemOpen
  \bibfield  {author} {\bibinfo {author} {\bibfnamefont {P.~M.}\ \bibnamefont
  {Morse}}\ and\ \bibinfo {author} {\bibfnamefont {H.}~\bibnamefont
  {Feshbach}},\ }\href@noop {} {\emph {\bibinfo {title} {Methods of theoretical
  physics}}}\ (\bibinfo  {publisher} {McGraw-Hill},\ \bibinfo {year}
  {1953})\BibitemShut {NoStop}%
\bibitem [{\citenamefont {Tan}\ and\ \citenamefont {Zhang}(2011)}]{tan2011}%
  \BibitemOpen
  \bibfield  {author} {\bibinfo {author} {\bibfnamefont {H.-T.}\ \bibnamefont
  {Tan}}\ and\ \bibinfo {author} {\bibfnamefont {W.-M.}\ \bibnamefont
  {Zhang}},\ }\href@noop {} {\bibfield  {journal} {\bibinfo  {journal}
  {Physical Review A}\ }\textbf {\bibinfo {volume} {83}},\ \bibinfo {pages}
  {032102} (\bibinfo {year} {2011})}\BibitemShut {NoStop}%
\end{thebibliography}


\newpage
\cleardoublepage
\newpage

\begin{center}
\textbf{\large SUPPLEMENTAL MATERIAL}
\end{center}

\setcounter{enumi}{1}
\setcounter{equation}{0}
\renewcommand{\theequation}{\Roman{enumi}.\arabic{equation}}

\setcounter{figure}{0}
\renewcommand{\thefigure}{\Roman{enumi}.\arabic{figure}}

\section{I. Normal Phase }
A normal mode with time-dependent frequency modulation [cf.~Eq.~\eqref{ham-nm} in the main text] can be generically described by  the Hamiltonian
for a parametric oscillator:
\begin{equation}\label{ham1}
H= \frac{p^2}{2m} + \frac12 [\omega_0^2 +\mu(t)] x^2\,.
\end{equation}
For a classical oscillator with sinusoidal modulation, the above Hamiltonian describes the well known Mathieu parametric oscillator \cite{{mclachlan1964,nayfehnonlinear}}. The quantum parametric Hamiltonian can be rewritten as  
\begin{align} \label{aham-1}
H& = [\omega_0 + \frac{\mu(t)}{2 \omega_0}] a^\dagger a + \frac{\mu(t)}{4 \omega_0} (a^2 + a^{\dagger 2})\nonumber\\
&\equiv \sqrt{\omega^2_0 + \mu(t)} {\bar a}^\dagger {\bar a}\,,
\end{align}
where the ladder operators $a, a^\dagger$ are defined with respect to the time independent model (i.e., $\mu(t)=0$).  

\begin{figure}[htbp]                   
\centering
\includegraphics[width=\columnwidth]{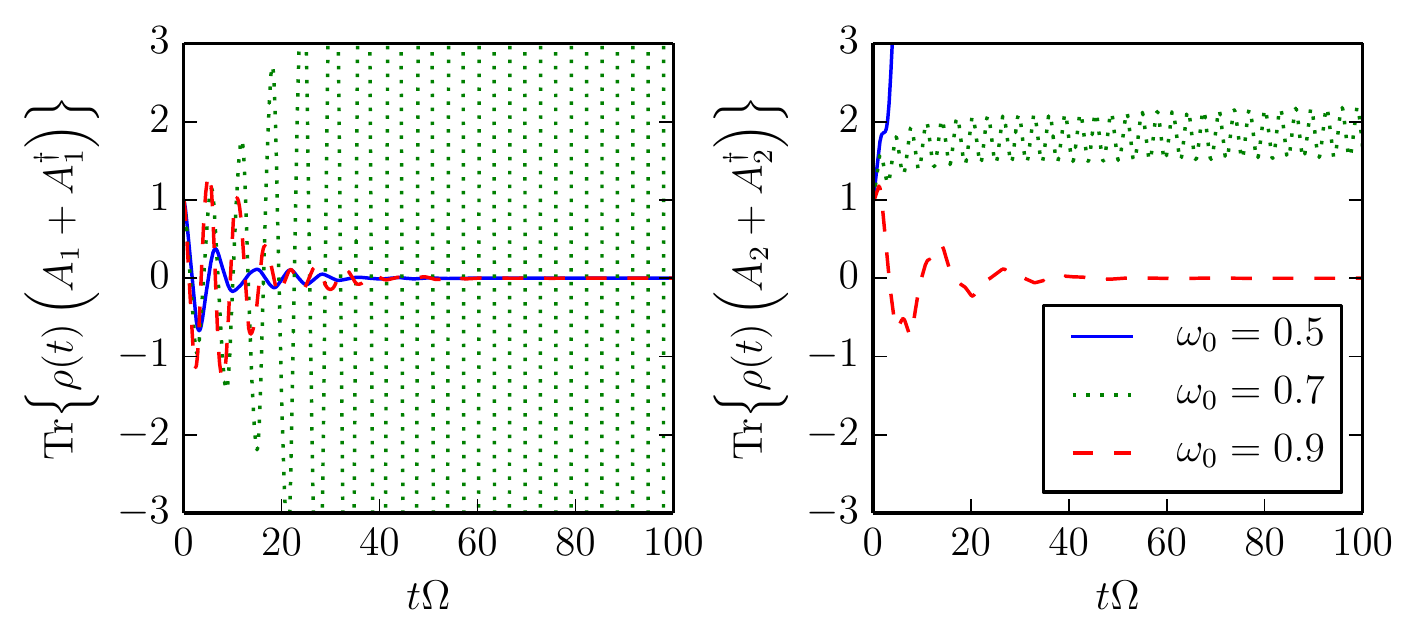} 
\caption{\label{gl_curves}Characteristic numerical time-integration plots of Eq.~\eqref{gl_eq} in different regions of the phase diagram. All curves are with $\lambda_0=0.4\Omega$, $\epsilon=0.5\Omega$, and $\kappa=\eta=0.1\Omega$. We see that for (i) $\omega_0=0.5\Omega$, mode $A_1$ is stable whereas $A_2$ is not, (ii) $\omega_0=0.7\Omega$, mode $A_1$ is unstable whereas $A_2$ is stable, and (iii) $\omega_0=0.9\Omega$, both modes $A_m$ are stable.}
\end{figure}

We consider a coupling to an external bath which is in the rotating wave approximation. Hence, the solutions to the Heisenberg equations of motion, in the presence of dissipation, for the operators $a$ and $a^\dagger$ take the form~\cite{tan2011}
\begin{eqnarray}
a(t) &=& G(t) a(0) + L^*(t) a^\dagger (0) + F(t)\,, \\
a^\dagger (t)&=& G^*(t) a^\dagger(0) + L(t) a (0)+ F^\dagger(t)\,,
\label{soln}
\end{eqnarray}
where $G$ and $L$ are time dependent functions obeying the initial conditions  $G(0)= G^*(0)=1$ and $L(0)=L^*(0)=0$.
$F$ is an operator term which stems from the dissipation and also depends on the functions $G$ and $L$.  It satisfies 
 the condition  $F(0)=F^\dagger(0)=0$. The functions $G,L$ obey the integro-differential equations
\begin{align} \label{int-diff1}
\resizebox{.85\hsize}{!}{$\displaystyle{\dot G}(t) = -i (\omega_0 + \frac{\mu(t)}{2 \omega_0}) G - i \frac{\mu(t)}{2 \omega_0} L  -\int_0^t ds K(t -s) G(s)\,,$}\\
\resizebox{.85\hsize}{!}{$\displaystyle{\dot L}(t)= i (\omega_0 + \frac{\mu(t)}{2 \omega_0}) L + i \frac{\mu(t)}{2 \omega_0} G -\int_0^t ds K(t -s) L(s)\,,$}
\label{int-diff2}
\end{align}
where  the dissipative kernel $K(t) = \int d\omega J(\omega) e^{-i \omega t}$ and  $J(\omega)$ is  the spectral density of
the dissipative bath.

\begin{figure}               
\centering
\includegraphics[width=\columnwidth]{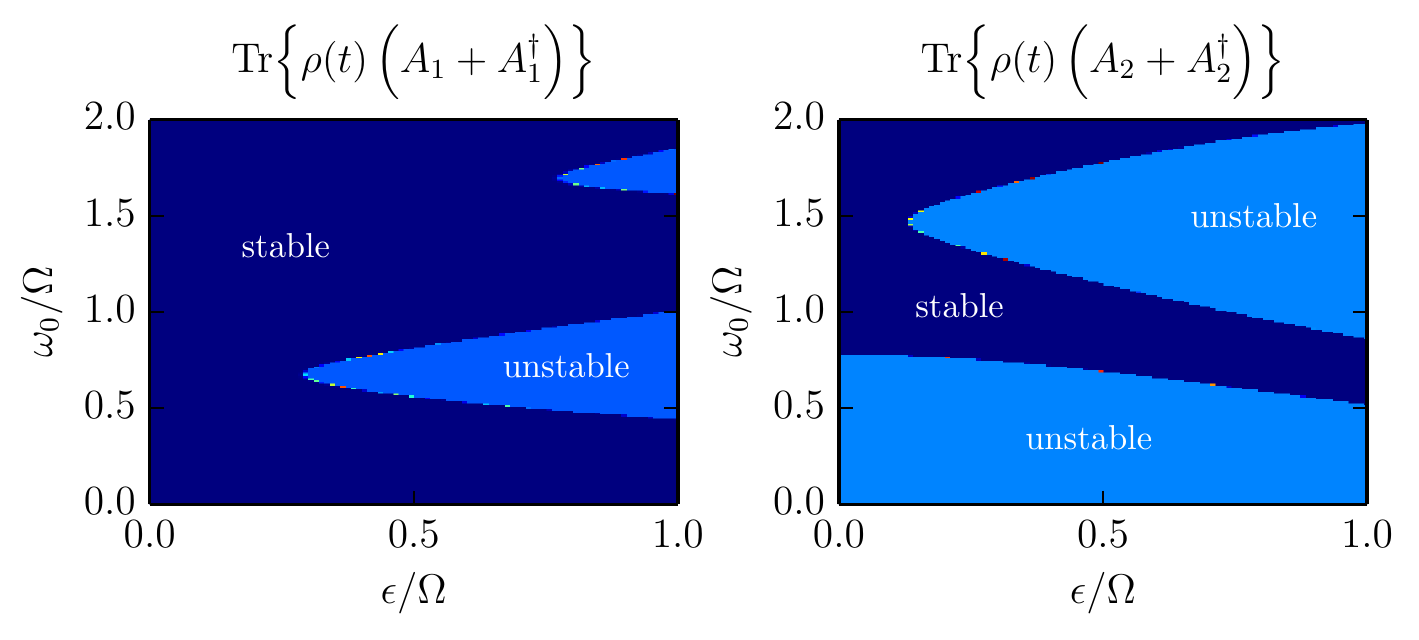} 
\caption{\label{gl_stability}The numerical stability diagram of the normal modes $A_m$ [cf.~Eq.~\eqref{ham-nm} in the main text] for $\lambda_0=0.4\Omega$ and $\kappa=\eta=0.1\Omega$. As the drive in the Dicke model affects each mode differently, we obtain two ``Arnold tongue'' stability diagrams that are shifted and scaled with respect to each other.  Superposing the zones where both modes are stable leads to the stability of the normal phase (NP) [cf.~Fig.~\ref{fig:diagram} in the main text].}
\end{figure}
 
These coupled first order integro-differential equations are rather difficult to solve  and numerical
solutions are needed. However, for standard Markovian dissipation, induced by cavity leakage in the rotating frame or coupling to an ohmic bath,  $K(t) = \gamma \delta(t)$ where $\gamma$ is the damping rate.  Substituting this in Eqs.~\eqref{int-diff1} and \eqref{int-diff2}, we see that they become a set of linear ODEs.
Observables and correlation functions can easily be obtained from these solutions. For example,
\begin{equation}
\langle a(t) \rangle = G(t) \langle a(0) \rangle + L^*(t) \langle a^\dagger(0) \rangle +  \langle F(t) \rangle\,,
\end{equation}
where the expectation values are with respect to the initial density matrix.
Assuming initial conditions such that $\langle F(t)\rangle =0$, which are expected for baths with no particular ordering, we  obtain
\begin{align}
\label{gl_eq}
\langle x(t) \rangle &={\rm Tr}\left\{\rho(t)\left(a+a^\dagger\right)\right\} \\
&= {\rm Re}[G(t)+L(t)] \langle x(0) \rangle -  {\rm Im}[G(t)+L(t)] \frac{\langle p(0) \rangle }{m \omega_0}\,.\nonumber
\end{align}
For arbitrary initial conditions, the stability of the oscillator is dictated by whether the pre-factors $\left[G(t)+L(t)\right]$ grow with time as one approaches the asymptotic state. For the parametric oscillator, we expect it to become exponentially unstable as the strength of the driving is increased \cite{zerbe1994}.  Choosing $\mu(t) = g \cos(2\Omega t)$, the zones of stability can be traced in the $\omega_0-g$ plane. The resulting stability diagram is the same as that for the classical Mathieu oscillators, which can also be extracted from the classical equations of motions [cf.~Eq.~\eqref{cla-eom} in the main text].

To obtain the full NP stability diagram of the Dicke model [see Fig~\ref{fig:diagram} in the main text], we study the stability of both normal modes $A_m$, $m=1,2$, with frequencies [cf.~Eq.~\eqref{ham-nm} in the main text]
\begin{eqnarray}
\Omega_1^2(t) &=& \omega_0^2 + 2 \lambda_0 \omega_0 + 2 \omega_0  \epsilon \cos(2\Omega t)\,, \\
\Omega_2^2(t) &=& \omega_0^2 - 2 \lambda_0 \omega_0 - 2 \omega_0  \epsilon \cos(2\Omega t)\,.
\end{eqnarray}
To simplify our calculation, we also assume that the baths the two modes couple to, have identical spectral densities~\cite{galve2010}. Though
relaxing this condition would lead to more  technical complexity, it should not have any nontrivial  physical consequence in the limit of
weak dissipation studied here.

We solve the corresponding Eqs.~(\ref{int-diff1})-(\ref{int-diff2})  and in Fig.~\ref{gl_curves}, we present characteristic numerical time-integration plots of Eq.~\eqref{gl_eq} for the normal modes $A_m$.    Repeating this procedure for different $\omega_0$ and $\epsilon$,  and  by  checking for converging/diverging solutions [cf.~Fig.~\ref{gl_curves}] we find the stability diagram for  both modes $A_m$, see Fig.~\ref{gl_stability}. 
The superposition of the two stability diagrams then yields the stability of the normal phase shown in Fig.~\ref{fig:diagram}.

\section{II. Mean-field analysis}
\setcounter{enumi}{2}

\setcounter{figure}{0}
\renewcommand{\thefigure}{\Roman{enumi}.\arabic{figure}}

\begin{figure}                   
\centering
\includegraphics[width=\columnwidth]{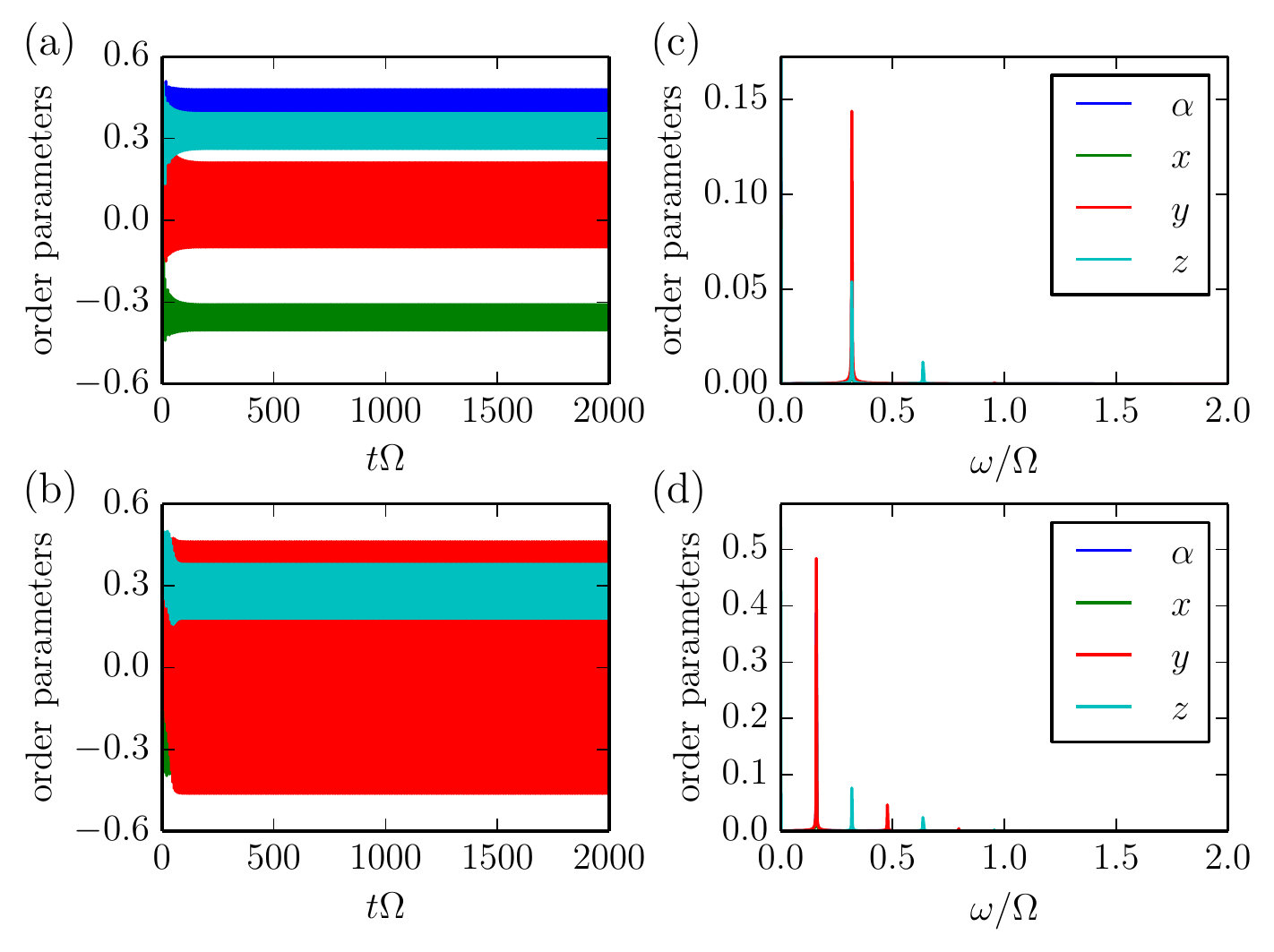} 
\caption{\label{mf_curves}Characteristic numerical analysis of the mean-field equations [cf.~Eqs.~\eqref{amean}-\eqref{ymean} in the main text] leading to the displayed phase diagram [cf.~Fig.~\ref{fig:diagram} in the main text]. All plots are with $\lambda_0=0.4\Omega$, $\epsilon=0.5\Omega$, and $\kappa=\eta=0.1\Omega$. (a) and (c) are characteristic numerical time-integration plots of the mean-field equations with $\omega_0=0.6\Omega$ and $\omega_0=0.65\Omega$, respectively. (b) and (d) are the corresponding Fast Fourier Transform (FFT). We see that in the super-radiant phase (SP) region [Figs.~(a) and (b)], all order parameters are oscillating around a non-zero mean with relatively small oscillations amplitudes. In the dynamical normal phase (D-NP) region [Figs.~(c) and (d)], apart from $z$, all order parameters have a zero mean value, but their oscillations are large taking the full length of the central spin in its $x$ and $y$ components.}
\end{figure}

In the main text, we obtained a set of coupled non-autonomous mean-field equations [cf.~Eqs.~\eqref{amean}-\eqref{ymean} in the main text] for the mean field parameters $\alpha, x  ~{\rm and} ~y$.  These equations were solved numerically with a variety of  ODE solvers. Characteristic numerical time-integration plots of the solutions to these equations  are shown in Figs.~\ref{mf_curves} (a) and (b).  Note that the  solutions converge to the  asymptotic regime  for times $t\Omega \sim 500$ and are oscillatory. The time scale for reaching the asymptotic regime varies with the parameters. Repeating this procedure as a function of $\omega_0$ and $\epsilon$, the time-average of the order parameters over the steady-state behaviour (the last one-third of the integrated time) is presented in Figs.~\ref{fig:diagram}(a) in the main text.  

We find that typically the solutions show sinusoidal oscillations characterized by the frequency of the parametric drive. However, 
in certain parameter regimes,  the order parameters  oscillate strongly  with a complex beat structure involving multiple frequencies. To analyze all these solutions in a systematic manner, we Fast Fourier Transform (FFT) the steady-state signals, see Figs.~\ref{mf_curves}(c) and (d). The largest amplitude of a finite frequency in such plots serves as a measure for the extent of the oscillation that the order paramaters undergo. This amplitude is plotted as a function of $\omega_0$ and $\epsilon$ in Figs.~\ref{fig:diagram}(b) in the main text.

\end{document}